\def\mpl{M_{\rm Pl}}
\def\Ob{\Omega_{\rm b}}
\def\Oc{\Omega_{\rm cdm}}
\def\Ok{\Omega_{ k}}
\def\Ol{\Omega_\Lambda}
\def\O5{\Omega_{r_c}}
\def\Om{\Omega_{ M}}

\def\ob{\omega_{\rm b}}
\def\od{\omega_{\rm d}}

\def\om{\omega_{\rm M}}

\def\d{{\bf d}}
\def\th{\hbox{\boldmath $\theta$}} 
\def\beq{\begin{eqnarray}}
\def\eeq{\end{eqnarray}}
\def\M{M_{(5)}}

\def\tmplbis{{\tilde{M}}_{\rm Pl}^2}
\def\tmpl{\tilde{M}_{\rm Pl}}

\documentstyle[12pt,psfig,epsfig,graphicx,psfrag]{article}

\begin{document}
\begin{flushright}
NYU-TH/01/08/03 \\
astro-ph/0201164\\
\end{flushright}
\vskip 1cm

\begin{center}
{\Large {\bf Supernovae, CMB,  and Gravitational Leakage into Extra Dimensions}}\\[1cm]
C\'edric Deffayet$^\dagger$\footnote{cjd2@physics.nyu.edu}, Susana J. Landau$^
{\ddagger \dagger}$\footnote{slandau@fcaglp.fcaglp.unlp.edu.ar},
Julien Raux$^*$\footnote{raux@in2p3.fr},Matias Zaldarriaga$^\dagger$\footnote{mz31@nyu.edu},\\
Pierre Astier$^*$\footnote{pierre.astier@in2p3.fr}.\\
$^\dagger${\it NYU Dept of physics, 4 Washington Place, New York, NY10003}\\
$^\ddagger${\it Observatorio Astron\'omico, Universidad Nacional de La Plata, Paseo del Bosque S/N, CP 1900 La Plata, Argentina.}\\
$^*${\it LPNHE, CNRS-IN2P3 and  Universit\'es Paris VI \& VII, Paris.}\\
\end{center}
\vskip 0.2cm

\noindent
\begin{center}
{\bf Abstract}
\end{center}

We discuss observational constraints coming from CMB and type Ia
supernovae, for the model of accelerated universe produced by gravitational
leakage into extra dimensions. Our fits indicate that the model is
currently in agreement with the data. We also give the equations
governing the evolution of cosmological perturbations. Future
observations will be able to severely constrain the model.

\vskip 1cm 

\section{Introduction}

Supernovae observations have recently provided evidence that the
expansion of the Universe is undergoing a late time acceleration
\cite{cc,SCP,Riess:2001gk}.  This acceleration can be explained in the
framework of standard cosmology by a non vanishing cosmological
constant. Although in agreement with current observations, such an
explanation exacerbates the usual cosmological constant problem
because it requires an explanation for its  very small, but non
zero, value.

One may wish to find alternative explanations for the acceleration,
and there are several proposals in the literature.  
Here we explore a scenario proposed  in \cite{Deffayet:2001uy,Fifth},  based
on the model of Dvali-Gabadadze-Porrati of brane-induced gravity \cite{DGP}.
 This proposal explains the observed late time
acceleration of the expansion of the Universe through a large scale
modification of gravity coming from ``leakage'' of gravity at large
scale into an extra-dimension, and without requiring a non vanishing
cosmological constant. The interesting point about this model 
from a phenomenological perspective is that it is a testable alternative 
to a cosmological constant model with the same number of parameters. 
This is in contrast to models of ``quintessence'' where the equation of
state of the new component becomes a free function that needs to be
constrained. 

In \cite{Fifth} it has been shown that the model was in qualitative
agreement with all known cosmological observations. The purpose of
this work is to go one step further and quantitatively confront the
model with observations of supernovae and the cosmic microwave
background (CMB).

The paper is organized as follows, in section \ref{Gravity} we discuss
the dynamics of the background metric of the universe in the
model. We first introduce in a few words the brane-induced gravity
model of Dvali-Gabadadze-Porrati  \cite{DGP} (see also \cite{DG,Dvali:2001gm,Dvali:2001gx}) which provides the framework
(subsection \ref{branind}). We then discuss the cosmological dynamics
for the accelerated solution considered in this paper (subsection
\ref{cosmdyn}). In the following, we confront the model with the Supernovae observations 
of
the Supernova Cosmology Project (SCP) \cite{SCP} (subsection  \ref{SNsect}) and CMB data (subsection  \ref{CMB}).

Our fits indicate that the model is currently in agreement with SNIa and small scale CMB data.
One can hope to discriminate the model from standard cosmology using future precision cosmological parameters measurements, but also maybe  modifications in the growth of large scale structures. 

\section{Model definition and Background dynamics} \label{Gravity}

In the following subsections we summarize the main features of the model under
consideration, study the dynamics of the background metric and
confront the predictions of the model to the supernovae observations. 

\subsection{Brane-Induced Gravity Models in a Few Words} \label{branind}

The brane-induced gravity models are a particular class of brane-world
models, which can be defined as models where our four dimensional (4D)
universe is considered to be a surface (called {\it brane}) embedded
into a higher dimensional {\it bulk} space-time. 

Brane-world models are
inspired by superstring-M theory, and can be regarded as some low
energy effective models of more fundamental underlying theories, but
have also interest on their own in providing new phenomenological
ideas.  We will only consider here the case where the bulk is five
dimensional (5D).  The brane embedding into the bulk is defined by the
coordinates $X^A(x^\mu)$ of the brane worldvolume (parametrized by
coordinates $x^\mu$) into the 5D space-time.  The dynamics of gravity
is governed by the usual 5D Einstein-Hilbert action 
\beq \label{5DEH}
S_{EH}~=~ {\M^3\over 2}~\int d^5X ~\sqrt
{|^{(5)}g|}~{^{(5)}R}\label{1}, 
\eeq 
where $\M$ denotes the 5D reduced Planck mass. The bulk metric
$^{(5)}{g}_{AB}$ induces through the embedding $X^A(x^\mu)$ a metric
$g_{\mu \nu}$ on the brane (called induced metric) defined by\footnote{In the following, we use upper case 
 Latin letters $A,B,...$ to denote 5D indices, Greek letters $\mu, \nu,...$ to denote indices parallel to the brane world volume, $5$  an index transverse to the brane,  and Latin letters $i,j,...$ to denote space-like indices parallel to the brane world volume}.  
\beq
g_{\mu \nu} =^{(5)} g_{AB} \partial_\mu X^A \partial_\nu X^B.  
\eeq 
In the above equation, we have
put an upper index $^{(5)}$ on quantities (e.g.  the 5D Ricci scalar
$^{(5)}R$ or the 5D metric $^{(5)}g_{AB}$) to distinguish them from
their 4D counterparts depending only on the induced metric (e.g. $R$ or $g_{\mu \nu}$).

In the brane-induced gravity models
\cite{DGP,DG,Dvali:2001gm,Dvali:2001gx,Dvali:2001ae}, the gravitational action
contains an other term $S_{eh}$, in addition to the 5D
Einstein-Hilbert term (\ref{5DEH}), given by
\beq \label{2}
S_{eh}~=~{\mpl^2\over 2}~\int_{brane} d^4x ~\sqrt {|g|}~{R}.
\eeq 
This term is the usual 4D Einstein-Hilbert term computed here on the
brane and with the induced metric, with  $\mpl$  a
mass parameter. The latter can be interpreted as the usual 4D reduced Planck
mass, from the calculation (see below) of the force between two static massive
sources on a flat brane and bulk background\footnote{We will not address
here the issue of the vDVZ discontinuity, see \cite{Deffayet:2001uk} and \cite{Lue:2001gc,Andrei}
for discussions of this issue.}.
The origin of $S_{eh}$ in brane world models is discussed in more details in 
\cite{DGP,DG,Dvali:2001gm}. It arises generically from quantum correction
coming from the coupling between bulk space-time and brane localized
matter fields when the conformal invariance of the brane theory is
broken (see e.g.\cite{Adler:1982ri}).
In the model at hand the dynamics of gravity is then governed by
the sum of the two kinetic terms $S_{EH}$ and $S_{eh}$.

As a consequence of the presence of the brane-induced term (\ref{2}),
one can show \cite{DGP} than the gravitational force experienced by
two static point like sources located on the brane is the usual 4D
gravitational $1/r^2$ force for distances smaller than the crossover
scale $r_c$ defined by 
\beq \label{rc} r_c = \frac{\mpl^2}{2
M_{(5)}^3}.  
\eeq 
For distances larger than $r_c$, on the other hand, the force turns to a
5D regime where it follows the 5D $1/r^3$ behavior. On scales smaller
than $M_{(5)}^{-1}$ one also expects modifications in the gravity law,
however for the parameter choice relevant to this work, the
modifications occur on scales much smaller than those
accessible by gravity experiments \cite{Dvali:2001gx}.  

This perturbative behavior has an exact parallel in cosmology, where
one can show \cite{Deffayet:2001uy} that,  for a $Z_2$
symmetric brane world (see
\cite{Dick:2001sc,Cordero:2001qd,Dick:2001np} for discussions of cases
where the $Z_2$ symmetry is relaxed),  the expansion of the
Universe is governed by the usual 4D Friedmann's equations whenever
the Hubble radius $H^{-1}$ is smaller than $r_c$, and enters into a
non conventional regime for larger Hubble radii.

In the following subsection we will discuss in greater detail this
cosmological evolution. At this point let us first say that an obvious
criteria that the model should reach in order to comply with the known
behavior of gravity at large distance, as well as with the observed
cosmology, is that $r_c$ should be made large enough. The more
stringent limit comes indeed from cosmology requiring $r_c$ to be of
the order of, or larger than, the today's Hubble radius $H_0^{-1}$.
When $r_c \sim H_0^{-1}$, one thus expects that cosmology is very close
to standard cosmology up to very late time, and in particular all
successes of standard cosmology such as BBN are left unchanged by this
choice of parameters. However the very recent evolution of the
universe is different.  Indeed, as will be reminded in more detail in
the next subsection, a particular class of solutions 
 shows a late time accelerated expansion
without the need for a non zero cosmological constant.  For values of
$r_c$ of order $H_0^{-1}$, as needed to fit the Supernovae
observations (see section \ref{SNsect}), one finds from Eq. (\ref{rc})
that $M_{(5)} \sim 10-100 $ MeV. Such a low value of the 5D Planck
mass is perfectly consistent with observations and high energy
experiments as shown in \cite{Dvali:2001gm, Dvali:2001gx}.  
Induced-gravity models have been shown to provide a framework
for realizing models with a very low quantum gravity scale without
conflicting with any experimental facts \cite{Dvali:2001gx}.

\subsection{Background cosmological dynamics}
\label{cosmdyn}

In the model considered here, the geometry of our 4D Universe is at
all time described by an ordinary FLRW space-time with a line element
of the form
\begin{eqnarray} \label{FLRW}
ds^2~ &=& g_{\mu \nu} dx^{\mu} dx ^\nu\\
&=& ~ -dt^2~ + ~a^2(t)~ dx^i~ dx^j ~\gamma_{ij},\\
&=&  -dt^2 + a^2(t)\left(dr^2 + S^2_k(r)d\psi^2 \right),
\end{eqnarray}
where $\psi$ are angular coordinates, $k=-1,0,1$ parametrizes the
brane world spatial curvature, and $S_k$ is given by
\begin{eqnarray}
S_k(r) = \left\{\begin{array}{ll} \mbox{sin } r & (k=1) \\
\mbox{sinh } r &(k=-1) \\ r &(k=0) \end{array} \right..
\end{eqnarray}
The cosmological standard observers are assumed, as usual, to be at
rest with respect to the comoving coordinates $x^i$. The only
difference with standard cosmology is in the dynamics of the metric
which is encoded into Friedmann-like equations different from to the
ordinary 4D ones.  For a given content of the universe, with total
energy density $\rho$ (and pressure $p$), the standard first
Friedmann's equation is now replaced by \cite{Deffayet:2001uy}
\begin{equation} \label{fried}
H^2 + \frac{k}{a^2} = \left(\sqrt{\frac{\rho}{3 {\mpl^2}}  +
 \frac{1}
{4 r_c^2}} + \frac{1}{2 r_c}\right)^2, 
\end{equation}
where 
\beq H \equiv \frac{1}{a} \frac{da}{dt},
\eeq 
is the Hubble parameter of our universe\footnote{ There is another set of solutions for a $Z_2$
symmetrical brane. Those were derived in
\cite{Deffayet:2001uy} and are not considered here.}. The
energy-momentum conservation equation, on the other hand, takes the
usual form \beq \label{cons} \dot{\rho} + 3H (p + \rho) = 0.  \eeq
Equations (\ref{fried}) and (\ref{cons}) are all what is needed to
characterize the cosmology we are interested in here.  They  
 lead to, 
\begin{equation} \label{H5D}
H^2(z) = H_0^2 \left\{ \Omega_k (1+z)^2 +
\left( \sqrt{\Omega_{r_c}} +
\sqrt{ \Omega_{r_c} + \sum_\alpha \Omega_\alpha(1+z)^{3(1+w_\alpha)} }
\right)^2 \right\},
\end{equation}
where $z$ is the redshift and
we have assumed that $\rho$ is given by the sum of the energy
densities $\rho_\alpha$ of different components (labeled by $\alpha$)
with constant equation of state parameters $w_\alpha$. 
The $\Omega$s for matter and curvature are defined in the usual way by
\begin{eqnarray} 
\label{omegafried}
\Omega_\alpha &\equiv& \frac{\rho^0_\alpha}{3 {\mpl}^2 H_0^2 a_0^{3(1+w_\alpha)}}~,
\\ 
 \Omega_k &\equiv& \frac{-k}{H_0^2 a_0^2}~,
\\
\end{eqnarray}
whereas $\Omega_{r_c}$ is given by 
\begin{eqnarray} 
\Omega_{r_c} &\equiv& \frac{1}{4 r_c^2 H_0^2}.
\end{eqnarray}
The normalization condition for the $\Omega$s, 
\begin{equation}\label{normalize}
 \Omega_k + \left( \sqrt{\Omega_{r_c}} + \sqrt{ \Omega_{r_c} +
  \sum_\alpha \Omega_\alpha} \right)^2 =1,
\end{equation}
differs from the usual relation $\Omega_k + \sum_\alpha \Omega_\alpha =1$.

Equation (\ref{fried}) implies that whenever $\rho /
\mpl^2$ is large compared to $1/r_c^2$ (or in other words, whenever
$H^{-1}$ is small with respect to $r_c$), the cosmological evolution
follows that of standard cosmology. In this case equation
(\ref{fried}) reduces to the standard Friedmann's
equation 
\beq H^2 + \frac{k}{a^2} = \frac{\rho}{3 \mpl^2}.  
\eeq 
When (and if) $\rho$ is driven to smaller values by the cosmic expansion,
the expansion of the Universe enters into a non conventional
phase and asymptotes to a de Sitter solution when $\rho$ becomes
negligible w.r.t.  $\mpl^2 / r_c^2$. One has a transition to an
accelerated expansion happening approximately when the Hubble radius
$H^{-1}$ crosses the threshold $r_c$. We would like to stress 
that this last accelerated phase is not triggered by a cosmological
constant (that can be consistently set to zero) but is due to the
presence of two kinetic terms for the graviton in the action. Namely,
bulk gravity sees the induced kinetic term on the brane (\ref{2}) as a
source term, and for an empty universe, there is a self-inflationary
solution\footnote{this solution is in a way the late time analog of
Starobinsky's first model of inflation where terms quadratic in the
Ricci tensor are sourcing similarly a self-inflationary solution \cite{Starobinsky:1980te}.} to
Einstein's equations to which a universe with decreasing energy will
asymptote. This solution acts as a late time attractor to early
standard cosmology.  

In the following we will then only consider a universe with a zero cosmological constant, and usual (dark, baryonic, ...) matter content.  
One can further notice that the above described  cosmology is
also exactly reproduced by standard cosmology with a dark energy
component with a $z$-dependent equation of state parameter
$w^{eff}_X(z)$.  For a universe containing only non relativistic matter, $w^{eff}_X(z)$ is given 
 by (see \cite{Fifth}) 
\beq w^{eff}_X(z) =
\frac{1}{\left(\sqrt{\frac{4 \Omega_{r_c}}{\Omega_M (1+z)^3} +
4}\right) \left( \sqrt{\frac{\Omega_{r_c}}{\Omega_M (1+z)^3}} +
\sqrt{\frac{\Omega_{r_c}}{\Omega_M (1+z)^3}+1} \right) } -1.
\eeq 
At large redshift $w^{eff}_X$ tends toward $-1/2$ reflecting the fact
that the dominant term in equation (\ref{H5D}), after matter and
curvature terms, redshift as $(1+z)^{3/2}$ at large $z$.  At low $z$,
however, $w^{eff}_X$ decreases toward an $(\Omega_k,
\Omega_M)$-dependent asymptotic value. For a flat universe, the latter
is simply given by\footnote{e.g. for $\Omega_M =
0.3$ and $k=0$, $w^{eff}_X$ at low $z$ tends toward $-0.77$.} $-1/(1+\Omega_M)$.  

In the following sections we give the results of fitting SNIa and
CMB observables with different cosmological parameters in the
framework of the cosmology defined by equations (\ref{fried}) and
(\ref{cons}). We will denote $\th$ a set of cosmological parameters
such as $\Omega_{r_c}$ or $\Omega_M$ characterizing a given cosmology.

\section{Confrontation with observations}
\subsection{Confrontation with supernovae observations} 
\label{SNsect}

We have fitted the supernovae data set from the SCP \cite {SCP} with
the luminosity distance calculated using Equation (\ref{H5D}). Because
the geometry of the Universe is given by usual FLRW (\ref{FLRW}) one
can use the standard formula for the luminosity distance $d_L$ as a
function of the redshift $z$, 
\begin{equation}\label{dL}
d_L = H_0^{-1} (1+z)\frac{ S_k \left( \sqrt{|\Omega_k|} d_C(z) \right)}{\sqrt{|\Omega_k|}},
\end{equation} 
with $d_C(z)$  defined by 
\beq
d_C(z) = \int_0^z H_0 \frac{dy}{H(y)}, 
\eeq 
and  $H(z)$ given by equation (\ref{H5D}). We then use this luminosity
distance to fit the data.  The fit is done using 4 free parameters:
the cosmological parameters $\th=(\Omega_M,\Omega_{r_c})$, the
intrinsic magnitude of the supernova $\cal{M}$, and a parameter
$\alpha$ related to the intrinsic luminosity-decline rate relation
(stretch factor s).  $\chi^2$ is given by
\begin{equation} \label{chi}
\chi^2(\th,\alpha,{\cal M})=\sum_{i=1}^{n}
{\frac{( {\cal M} + \alpha (1-s_i)+5~log_{10}
(d_L(\th,z_i))-m_i)^2}{\sigma_i^2}}
\end{equation}

\begin{figure}
\psfrag{z}{{\footnotesize Redshift}} \psfrag{m}{{\footnotesize $\Delta
m$}} \epsfig{file=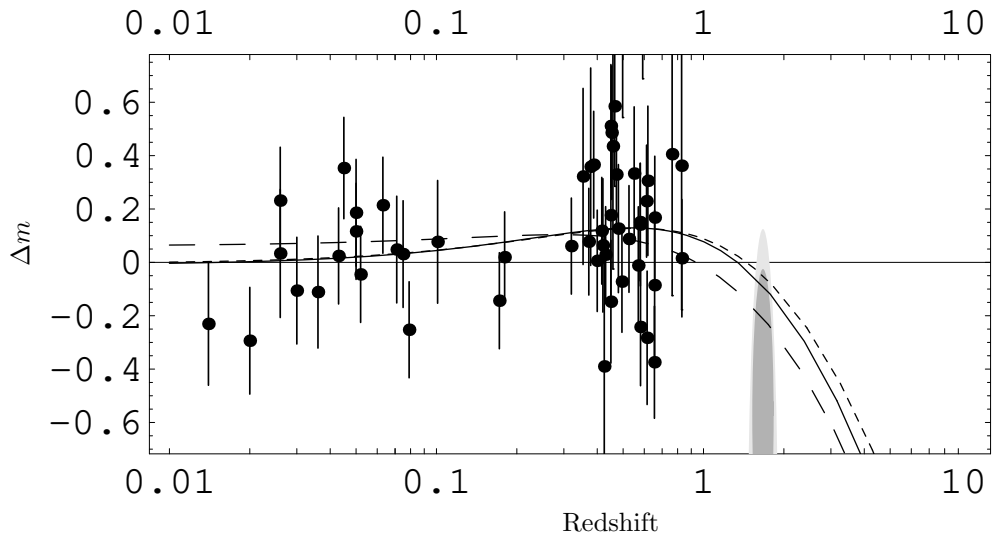, width=1.0\textwidth}
\caption{Magnitude vs redshift diagram for the SNIa data of reference
\cite{SCP} used in this paper. All
magnitude are plotted respectively to an empty universe ($\Omega_M=0$
and $\Omega_\Lambda=0$). Over-plotted are three different flat
cosmological models: the best fit flat model in standard cosmology
(with $\Omega_M=0.28$ and $\Omega_\Lambda=0.72$, solid line), in
the gravitational leakage cosmology (with $\Omega_M=0.18$, dotted line) 
and a flat model in the gravitational
leakage cosmology with $\Omega_M=0.3$ (dashed line).  We also show 
two approximate confidence level interval for the $z=1.7$ supernova of
Ref.  \cite{Riess:2001gk}, the outer light-gray surface represents
the $95 \% $ confidence interval, the inner dark gray
surface represents approximately the $68 \%$ confidence interval.
This last supernova was not included in the fit.
The values of $\alpha$ (related to the stretch factor) and ${\cal M}$ (intrinsic
magnitude) have been fitted independently for all the models.  The
data are plotted here with  $\alpha = 0.6$.}
\label{sndata}
\end{figure}

\begin{figure}[htb]
        \epsfig{file=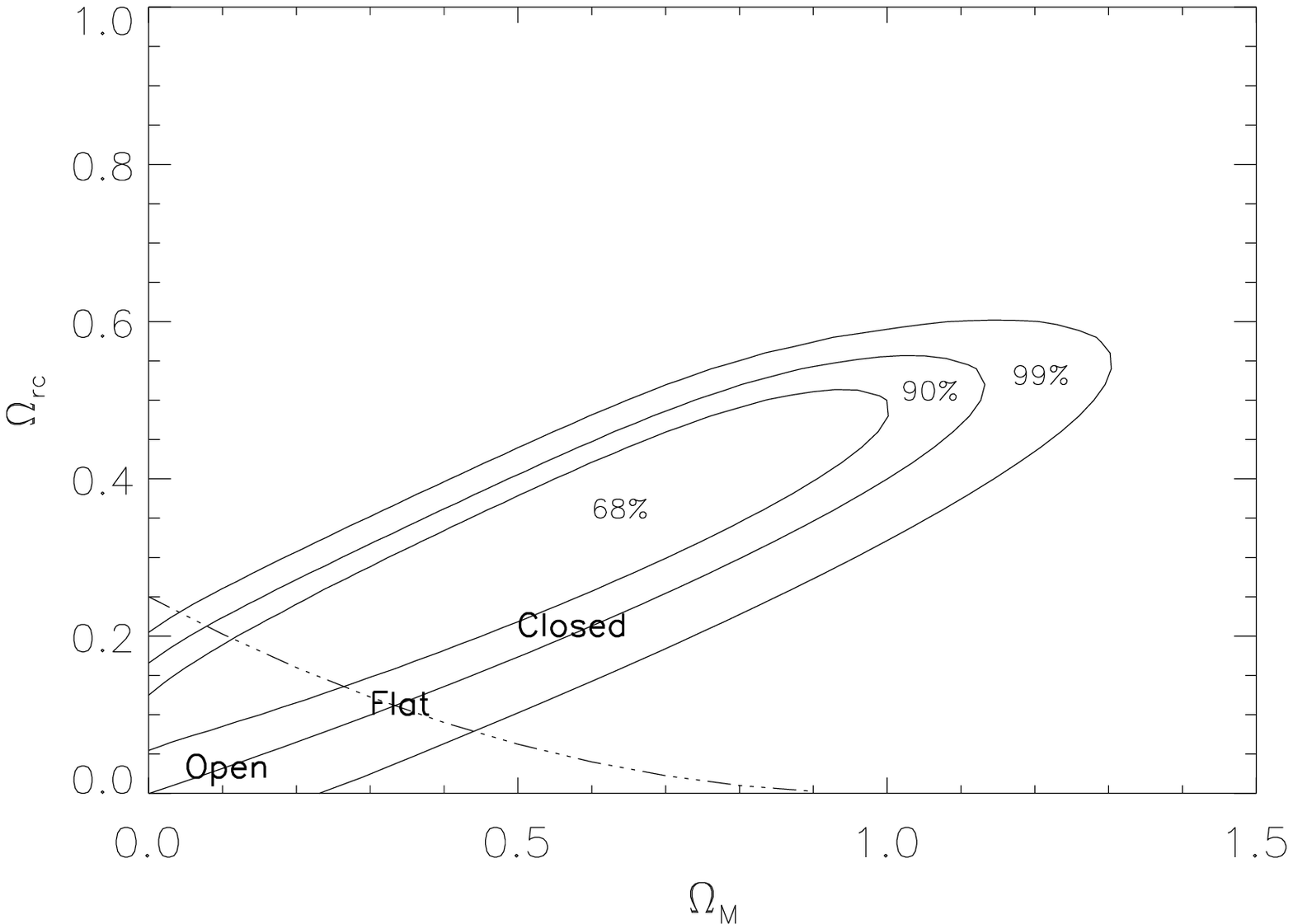, width=1.0\textwidth}
  \caption{68.3\%, 90\% and 99\% confidence regions for
    $(\Omega_M,\Omega_{r_c})$ in gravitational leakage scenario,
    assuming no priors knowledge of $\alpha$ and $\cal{M}$. \label{fig1}}
\end{figure}

The data set consisting of 54 supernovae (18 nearby ones and 36 at
high redshift) is shown in figure \ref{sndata}.  Since we assume no
prior knowledge of the parameters and as we are not interested by
$\alpha$ and $\cal{M}$, we have to marginalize over them. We do this
in a Bayesian framework assuming flat priors and Gaussian errors.
These integrations can be carried out analytically, as shown in
\cite{Goliath}. We quote the results in appendix \ref{appA}. We have
then computed confidence contours for the models
$\th=(\Omega_M,\Omega_{r_c})$ model with no prior on the
cosmology. These contours are plotted in fig \ref{fig1}.

Assuming a spatially flat space-time, one is only left with one free
parameter (after integration over $\cal{M}$ and $\alpha$),
e.g. $\Omega_M$. $\Omega_{r_c}$ is then given by the normalization
condition (\ref{normalize})
\begin{equation} \label{flat5}
\Omega_{r_c} = \left(\frac{1-\Omega_M}{2}\right)^2
\mbox{, $\Omega_{r_c} <1$ and $\Omega_M<1$}.
\end{equation} The results of the
$\chi^2$ minimization gives for a flat universe (one sigma levels)
\begin{eqnarray}
\Omega_M=0.18^{+0.07}_{-0.06} \quad \mbox{or} \quad  \Omega_{r_c}=0.17_{-0.02}^{+0.03},
\label{rcest}
\end{eqnarray}
with  $\chi^2=57.96$, for 52 (54 SNe - 2 parameters) degrees of freedom
\footnote{These numerical results are in agreement with the fit done
in \cite{Avelino:2001qh}.  We however disagree with the conclusions of
that work as will be discussed later (see also Ref.
\cite{Deffayet:2001xs} for a discussion of this paper). Note in
particular that, contrarily to the claims made in
\cite{Avelino:2001qh}, the $z=1.7$ supernova of Ref.
\cite{Riess:2001gk}, is fitted as well by the model considered in this
paper, or by standard cosmology with a cosmological constant(see
figure \ref{sndata}).}  This best fit model is shown in figure
\ref{sndata}.  Equation (\ref{rcest}) leads to an estimate for
$r_c$ in terms of the Hubble radius $H_0^{-1}$ given by \beq
\label{rcest2} r_c = 1.21^{+0.09}_{-0.09} H_0^{-1}.  \eeq

\subsection{Confrontation with CMB Observations}
Another set of cosmological observables which has recently been
measured with great precision is the CMB temperature power
spectrum. In this subsection we would like to compare the predictions of the
model considered in this paper to the results of these 
observations. 

For this purpose, we used a modified CMBFAST \cite{cmbfast} replacing
the first Friedmann's equation by equation (\ref{fried}).  The
equations for the growth of cosmological perturbations were kept the
same as in usual cosmology (except for the background evolution).  As
is discussed qualitatively in appendix B,  this is  justified for the
small scale perturbations and for processes happening early enough in
the history of the Universe.  On the other hand, one can expect 
deviations from the standard picture at large
scale (and late time) where (and when) the effect of the extra
dimension began to be felt.  This concerns scales of order of today's
Hubble radius and processes happening in the late history of the
universe. A more refined discussion of this, which involves the integration of bulk 
equations of motions for perturbations, is left for future work
\cite{prep}.
 
\label{pertCMB}
\label{3.1}
\label{CMB}

We explored the six-dimensional parameter space, $\th= (\Ok, \O5, \od,
\ob, n, A)$, where $\od=\Oc h^2,\ \ob=\Ob h^2$ and $A$ and $n$ are the
amplitude and slope of the primordial spectrum of perturbations. We
used a Markov chain method to explore the likelihood in this parameter
space. When it has converged the method produces a chain of models
that are sampled from the probability distribution of $\th$.  The
details of our procedure are given in  appendix \ref{appB}.

Figure \ref{marg1} shows the probability distribution for each of
the six parameters obtained. As expected the CMB data prefers
spatially flat models. Figure \ref{OmO5} shows the results of our
analysis in the $\Omega_{M} - \Omega_{r_c}$ plane. The shaded 
region was drawn to contain approximately 95 \% of the models in our
chain, the line marks the location of spatially flat
models. The constraint on $\O5$ is coming mainly from the position of
the acoustic peaks so there is a natural degeneracy in the $\Om - \O5$
plane which is apparent in the plot.

\begin{figure}[ht]
\centerline{\epsfxsize=9cm\epsffile{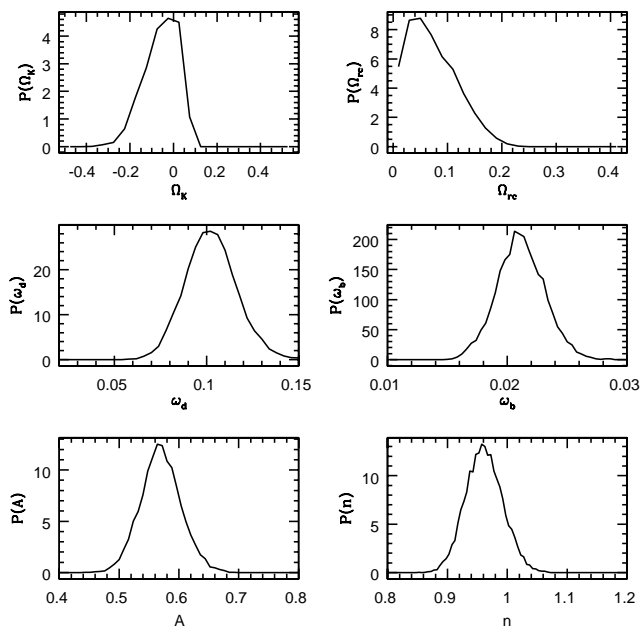}}
\caption{Marginal distribution for each of the 6 parameters used.}
\label{marg1}
\end{figure}

\begin{figure}[ht]
\centerline{\epsfxsize=9cm\epsffile{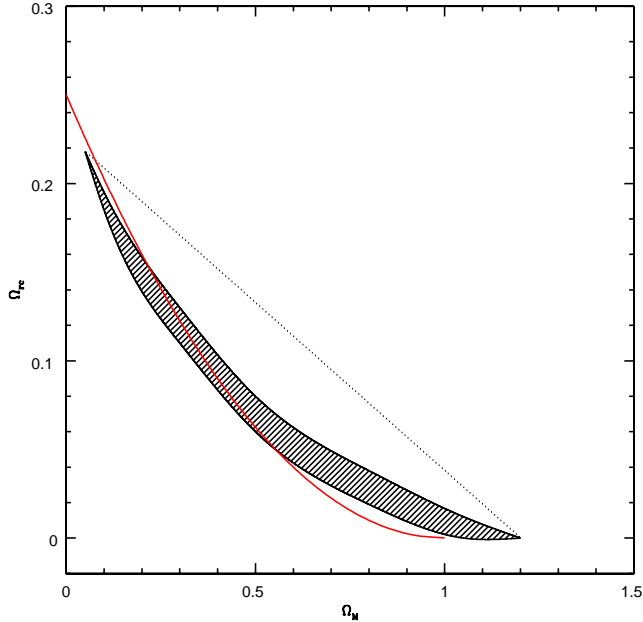}}
\caption{Allowed region in the $\Om - \O5$ plane (shaded). The line
shows the location of spatially flat models. The shaded region was
drawn to contain approximately 95 \% of the models in our chain.}
\label{OmO5}
\end{figure}

\begin{figure}[ht]
\centerline{\epsfxsize=9cm\epsffile{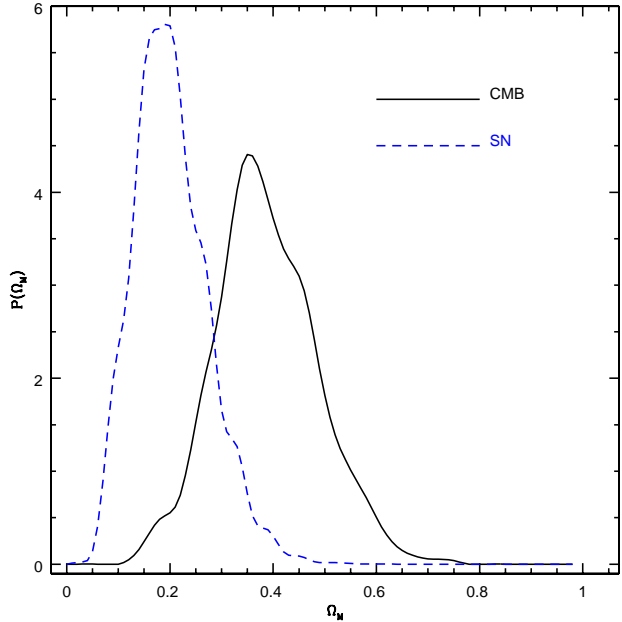}}
\caption{Marginal distributions for $\Om$ under the assumption that the universe
is spatially flat. The solid line shows the results from CMB and the
dashed line from the SN.}
\label{marg2}
\end{figure}

The probability distribution for $\Ok$ shown in figure \ref{marg1}
peaks around $\Ok=0$, a spatially flat universe. Thus it is natural to
further restrict ourselves to flat universes which we can do by
considering only samples in our chain with negligible curvature. The
probability distribution for $\Om$ under this assumption is shown in
figure \ref{marg2}.

Figure \ref{marg2} shows that a model with $\Om=0.3$ provides a good
fit to both SN and CMB data. It should be noted however that the CMB
prefers a slightly larger value of $\Om$ than the SN, although both
ranges overlap at 1$\sigma$. In turn, the CMB can constrain the
physical densities in matter quite accurately, $\om+\ob\approx
0.12$. This constraint translates $\Om=0.3$ to a Hubble constant
$h\approx 0.63$ in good agreement with direct measurements, {\it e.g}.
$h=0.72\pm0.08$ from the HST key project \cite{keyproj}.

\begin{figure}[ht]
\centerline{\epsfxsize=9cm\epsffile{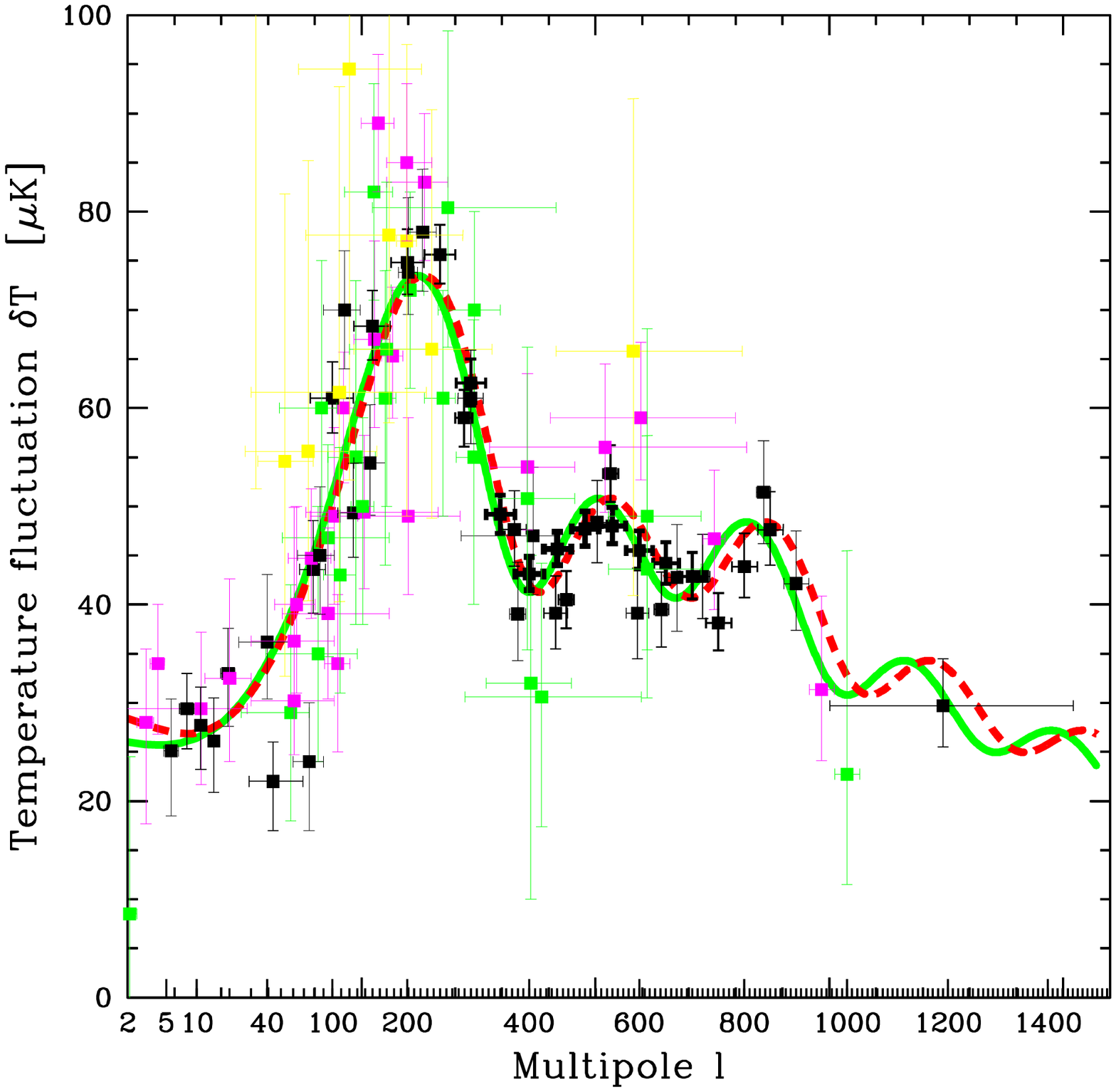}}
\caption{Model predictions and current CMB data. The solid curve curve
is for a model with $\th = (\Ok, \O5, \od, $ $ \ob, n, A) =
(0,0.1225,0.1,0.02,0.96,0.57)$ while the dash curve is for standard
cosmology with a cosmological
constant $\Ol=0.7$ (other parameters were kept the same).}
\label{cls}
\end{figure}

In figure \ref{cls} we show what we could call our ``concordance''
model, $\th = (\Ok, \O5, \od, $ $ \ob, n, A) =
(0,0.1225,0.1,0.02,0.96,0.57)$ which has $\Om=0.3$ and
$\chi^2\approx140$ for the full data set (135 data points).  For
reference we also show the power spectra for a standard cosmology
model with $\Ol=0.7$, a model that has all other parameters the same
but where the acceleration is produced by a cosmological
constant. Both models provide an acceptable fit to the current data,
but because there angular diameter distance to the last scattering
surface differs by approximately $4\%$ they should be easily
distinguishable by future generation of CMB experiments. The
difference at low multipoles should be regarded with care because on
this very large scales the physical effects we ignored could be
relevant (see appendix \ref{scalar}).

\section{Conclusions}

The fits done in this work show that the model of accelerated universe
through gravitational leakage into extra dimension of Ref.
\cite{Deffayet:2001uy,Fifth} is in current agreement with SNIa and CMB
data.  The degeneracies in parameters estimations using one data set
(e.g. CMB) can be partially lifted using the other (e.g. SNIa) as in
standard cosmology. The Supernovae data prefer a slightly lower value
of $\Omega_M$ ($\Omega_M = 0.18^{+0.07}_{-0.06}$) than the CMB for a
flat universe, however a concordance model with $(\Ok, \O5, \od, $ $
\ob, n, A) = (0,0.1225,0.1,0.02,0.96,0.57)$ which has $\Om=0.3$ (and
$\chi^2\approx140$ for the full data set (135 data points)) provide a
good fit to both sets, all the more as we have not included systematic
errors in our parameter estimations.  For this model the crossover
distance between 4D and 5D gravity is given by $r_c \sim 1.4~H_0^{-1}$.

We have also given the equation of evolution for cosmological
perturbations.  Those equations were used to justify the approximation
we made to compute cosmological perturbations, namely we used standard
four dimensional evolutions equations over a background with a scale
factor given by the accelerated solution given in \cite{Fifth}. This
is justified for small scale CMB anistotropies (scale smaller than the
crossover scale $r_c$). From those equations, and the known behavior
of gravity in the model at hand, one can also expect modifications in
the growth of large scale structure. This could potentially lead to a
way to discriminate between standard cosmology and the model
considered in this work, and is left for future investigation.

We want to end by noting that the model under consideration is very
predictive in the sense that future observations have the potential
to rule it out. In contrast to quintessence models, this model has the
same number of free parameters as the usual LCDM model. With the
advent of new precision cosmological measurements such as new SNIa
observations, CMB measurements, ongoing galaxy surveys such as Sloan
and 2dF, weak lensing surveys, etc. it should be possible to test the
model very accurately (for a recent summary of how different
observations will constrain the matter content of the universe see
\cite{Tegmark:met} and references therein).

\section*{Acknowledgments}
We thank Gia Dvali, Andrei Gruzinov, Arthur Lue, Roman Scoccimarro for
useful discussions.  The work of C.D. is sponsored in part by NSF
Award PHY 9803174, and by David and Lucile Packard Foundation
Fellowship 99-1462. MZ is supported by the David and Lucile Packard
Foundation and NSF grant AST-0098506 and NSF grant PHY-0116590

\appendix

\section { Marginalization for Supernovae}
\label{appA}
Following \cite{Goliath} the $\chi^2$ defined by equation (\ref{chi}) can be integrated analytically
over $\cal{M}$ and $\alpha$ to yield  
\begin{eqnarray}
    \chi^2_{{\alpha}{\rm-int}}(\th)&=&
    -2\ln\left[\int_{-\infty}^\infty\,d{\alpha}
    \exp\left(-\frac{1}{2}\chi^2_{{\cal M}{\rm-int}}(\th, 
    {\alpha})\right)\right]\\
    &=&A'-\frac{B'^2}{C'}-\frac{\left(F-\frac{B'E}{C'}\right)^2}{D-\frac{E^2}{C'}},\\
    A'&=&\sum_{i=1}^n\frac{\left(5\log_{10}\left[d_L(\th,z_i)\right]
      - m_i\right)^2}{\sigma_i^2} , \\
    B'&=&\sum_{i=1}^n\frac{5\log_{10}\left[d_L(\th,z_i)\right]
      - m_i}{\sigma_i^2} , \\
    C'&=&\sum_{i=1}^n\frac{1}{\sigma_i^2} ,\\
    D&=&\sum_{i=1}^n\frac{(1-s_i)^2}{\alpha_i^2} , \\
    E&=&\sum_{i=1}^n\frac{(1-s_i)}{\alpha_i^2} , \\
    F&=&\sum_{i=1}^n\frac{(5\log_{10}\left[d_L(\th,z_i)\right]-
      m_i)(1-s_i)}{\sigma_i^2} ,
  \end{eqnarray}
where   $\chi^2_{{\cal M}{\rm-int}}(\th,\alpha)$ is defined by 
\beq
 \chi^2_{{\cal M}{\rm-int}}(\th,\alpha)&=&
    -2\ln\left[\int_{-\infty}^\infty\,d{\cal M}
    \exp\left(-\frac{1}{2}\chi^2(\th, {\cal M}, \alpha)\right)\right]
\eeq

\section{Dynamics of scalar cosmological perturbations}
\label{scalar}
We briefly summarize here the equations governing the cosmological
perturbations in the model at hand. These equations will be derived and discussed in more
details elsewhere, and are only given here for the case of a flat universe.

Our starting point is an equation derived in \cite{Shiromizu:2000wj},
relating the 4D Einstein's tensor $G_{\mu \nu}$ to a tensor $\Pi_{\mu
\nu}$ quadratic in whatever source $\tilde{T}_{\mu \nu}$ of 5D
Einstein's equations is localized on the brane, and a traceless tensor
${\cal E}_{\mu \nu}$ defined in terms of the 5D bulk Weyl
tensor. The corresponding equation reads \beq \label{shiro} G_{\mu
\nu} = \frac{1}{M_{(5)}^6} \Pi_{\mu \nu} - {\cal E}_{\mu \nu}, \eeq
with $\Pi_{\mu \nu}$ given by \beq \Pi_{\mu \nu} &=& -\frac{1}{4}
\tilde{T}_{\mu \alpha} \tilde{T}^{\alpha}_\nu + \frac{1}{12}
\tilde{T}\tilde{T}_{\mu \nu} + \frac{1}{8} \tilde{T}_{\alpha \beta}
\tilde{T}^{\alpha \beta}g_{\mu \nu} - \frac{1}{24} \tilde{T}^2 g_{\mu
\nu}, \eeq and ${\cal E}_{\mu \nu}$ is defined by 
\beq \label{Weyldef}
{\cal E}_{\mu \nu} = C^5_{\quad \mu 5 \nu} 
\eeq from the bulk Weyl's
tensor\footnote{we have chosen here implicitly a
Gaussian normal coordinate with respect to the brane} $C^A_{\quad BCD}$. In our case $\tilde{T}_{\mu \nu}$
is given by
\begin{equation}
\tilde{T}_{\mu \nu} = T_{\mu \nu} -   \mpl^2   G_{\mu \nu},
\end{equation}
where $T_{\mu \nu}$ is the brane energy momentum tensor  and 
$G_{\mu \nu}$ is the 4D Einstein's tensor.  $T_{\mu \nu}$ 
is conserved with respect to the 4D metric on the brane, so that one
has 
\beq \label{conserve} D_{\mu} T^{\mu}_{\nu} &=& 0,\\ D_{\mu}
\tilde{T}^{\mu}_{\nu} &=& 0, 
\eeq
 where $D_\mu$ denotes the covariant
derivative compatible with the 4D metric on the brane, and the last
equality follows from Bianchi identities.  Equations (\ref{shiro}) and
(\ref{conserve}) lead to the background equation of motion
(\ref{H5D}), once one knows the background expression for ${\cal
E}_{\mu \nu}$. In the cosmological case, ${\cal E}_{\mu \nu}$ is in
general given by some version of Birkhoff's theorem
\cite{Kraus:1999it,Binetruy:2000hy,Flanagan:2000cu,Mukohyama:2000wi,Bonjour:2000ca}.
We have assumed for simplicity in (\ref{H5D}) that it vanishes in the
background, in which case the 5 dimensional space-time is simply a
Minkowski space-time\footnote{As far as the background is concerned a
non vanishing $\cal E_{\mu \nu}$ is manifesting itself as a radiation
component into the Friedmann's equations; see
Ref. \cite{Deffayet:2001uy} where the background equations are given
in full generality.}.

We now derive from equation(\ref{shiro}), the evolution equations
for the cosmological perturbations. We write 
\beq
G^\mu_{\nu} &=& ^{B}G^\mu_{\nu}+ \delta G^\mu_{\nu}, \\ T^{\mu}_{\nu}
&=& ^B T^{\mu}_{\nu} + \delta T^{\mu}_{\nu}, \\ {\cal E}^{\mu}_{\nu}
&=& \delta {\cal E}^{\mu}_{\nu}, 
\eeq 
where the superscript $^B$ stands for the background value of the
corresponding tensor component.  We define then the scalar
perturbations in energy density, $\delta \rho$, momentum, $\delta q$,
pressure, $\delta P$, and anisotropic stress, $\delta \pi$, for
ordinary matter as 
\beq \delta T^0_0 &=& -\delta \rho, \\ \delta T^0_i
&=& \nabla_i \delta q, \\ \delta T^i_j &=& \delta P \delta^i_j
+\left(\nabla^i \nabla_j - \frac{1}{3} \delta^i_j \nabla^2\right)
\delta \pi, 
\eeq 
where $\nabla_i$ is the covariant derivative adapted to the background
spatial metric $\gamma_{ij}$ parallel to the brane.  We also define
similar quantities for the{\it Weyl's fluid}, following
\cite{Langlois:2001ph,Langlois:2001iu,Langlois:2000ia,Bridgman:2001mc}
\beq \delta {\cal E}^0_0 &=& \frac{1}{\mpl ^2} \delta \rho_{\cal E},
\\ \delta {\cal E}^0_i &=& -\frac{1}{\mpl^2} \nabla_i \delta q_{\cal
E},\\ \delta {\cal E}^i_j &=&-\frac{1}{\mpl^2} \left(\delta P_{\cal E}
\delta^i_j + \left(\nabla^i \nabla_j - \frac{1}{3} \delta^i_j
\nabla^2\right) \delta \pi_{\cal E} \right).  
\eeq 
The {\it Weyl's fluid} is related to the perturbation of the bulk
Weyl's tensor (gravitational waves in the bulk) through equation
(\ref{Weyldef}).

Other useful quantities are the trace, $\delta G_T$, and traceless
traceless part, $\delta G_{TF}$, of $\delta G^i_j$ defined by 
\beq
\delta G^i_j = \delta G_T \delta^i_j + \left(\nabla^i \nabla_j -
\frac{1}{3} \delta^i_j \nabla^2\right) \delta G_{TF}.  
\eeq 
After some algebra, one gets then from equations (\ref{shiro}) the
perturbed Einstein's tensors over the background
(\ref{cons})-(\ref{H5D})
\beq
\label{perteq00} \delta G^0_0 \left(1 - \frac{1}{2 H r_c} \right) &=&
-\frac{1}{\mpl ^2} \left( \delta \rho - \frac{\delta \rho_{\cal E}}{2
H r_c} \right),\\ \delta G^0_i \left(1 - \frac{1}{2 H r_c} \right)
&=& \frac{1}{\mpl ^2} \left( \nabla_i \delta q - \frac{1}{2 H r_c} \nabla_i \delta q_{\cal E} \right)\label{perteq0i},\\
\label{perteqTF}
\delta G_{TF} \left(1-\frac{H}{ r_c (\dot{H}+2H^2)}\right) &=&
\frac{1}{\mpl ^2} \left( \delta \pi - \delta \pi_{\cal E} \frac{H}{
r_c (\dot{H}+2H^2)}\right),\\ \label{perteqT} \delta G_{T}
\left(1-\frac{1}{2 H r_c} \right) &=&\frac{1}{\mpl ^2} \left( \delta P
- \frac{1}{2 H r_c} \delta P_{\cal E} + \frac{\dot H}{3H^2}
\frac{\delta \rho - \delta \rho_{\cal E}}{2 H r_c-1} \right).  
\eeq
These equations replace the perturbed Einstein's
equations of ordinary cosmology (see e.g. \cite{Mukhanov:1992me} for a
review).

One can derive from (\ref{conserve}) the usual conservation equations
for the matter perturbations.  As far as the Weyl's fluid is
concerned, by taking the covariant derivative of
equation (\ref{shiro}) and using (\ref{conserve}) one can show 
that the Weyl's fluid energy density $\delta \rho_{\cal E}$ 
is conserved but that the
Weyl's fluid momentum $\delta q_{\cal E}$ in general is not 
\cite{Langlois:2001iu}. Moreover one does not have an evolution
equation for the Weyl's fluid anisotropic stress $\delta \pi_{\cal
E}$. This means that the system of equations for cosmological
perturbations do not close on the brane, and one needs to solve the
equations of motion for gravitational waves in the bulk (see
\cite{Langlois:2001iu}). On large scales, however, the usual adiabatic curvature
perturbation on hypersurfaces of uniform (ordinary or Weyl) matter
density is conserved \cite{Langlois:2001iu}, since it is a mere
consequence of the conservation of the energy density perturbation
\cite{Wands:2000dp}.  However one still cannot compute the Sachs Wolf
effect because of the lack of knowledge of $\delta \pi_{\cal E} \cite{Langlois:2001iu}$.

Let us make here some simple remarks. In the formalism used so far,
the deviation from usual 4D cosmological perturbations equations can
be separated into two different parts.

 We first note that the direct coupling between ordinary matter and
gravitational perturbations is $H$ dependent, for example one can
rewrite equation (\ref{perteq00}) as \beq \label{dG00} \delta G^0_0 =
-\frac{1}{\tmplbis} \left( \delta \rho - \frac{\delta \rho_{\cal E}}{2
H r_c} \right) \eeq with the effective direct gravitational coupling
between matter and gravity given by \beq \tmplbis = \mpl ^2 \left(1 -
\frac{1}{2 H r_c} \right).  \eeq One can check that this coupling is
never negative for the late time accelerated solution considered in
this paper, since one always has $H r_c \ge 1$. Moreover in the early time of the Universe
(whenever $H r_c \gg 1$), one has $\tmpl \sim \mpl$ so that one can
consistently ignore this effect at least up to last scattering (in
contrast to what is happening in usual brane cosmology), which is all
what matters as far as CMB is concerned.  At the epoch of last
scattering, for example, $\tmpl$ coincides with $\mpl$ within a part
per thousand.  However, $H r_c $ becomes of order unity at late time (see
equation (\ref{rcest})), and one can be concerned that this can have
dramatic effects on large scale structure formation\footnote{This
could also have potentially observable signature through standard
tests of gravity.  We however expect that when one looks at
fluctuations over a given background,  a local curvature scale $l^{-1}$
 should typically replace $H^{-1}$ in the above equations so that
we do not expect that the $l$ dependence of $ \tmpl$ could have
observable effects on systems where the curvature is much greater
than today's $H_0$ ($\sim r_c^{-1}$).
This issue is likely to be related to the disappearance of the vDVZ discontinuity (see \cite{Deffayet:2001uk,Lue:2001gc,Andrei}) and will be discussed elsewhere.}. To be consistent one should
also consider in this regime the effects of the Weyl's fluid source
terms in the left hand side of the perturbed Einstein's equations
(\ref{perteq00}-\ref{perteqT}) (as well as possible non linear corrections, see \cite{Deffayet:2001uk}) , and this can only be done properly
solving for the bulk equations of motions for perturbations.  With the
formalism used so far, those source terms are the other manifestation
of the extra dimension that we would like to discuss now qualitatively
as far as the CMB is concerned. 

One note, that those source terms are suppressed with respect to their
ordinary matter counterparts by a factor $H r_c$, in contrast to what
happens usually for brane cosmology (see e.g \cite{Bridgman:2001mc}).
This support the fact that when $H r_c \rightarrow \infty$ the theory
looks more and more 4 dimensional. One can then start with initial
conditions for cosmological perturbations, say after inflation, which
are the one provided by standard 4D cosmology, and set initially all
the Weyl's perturbations to zero. The brane perturbations will then
feed up non zero perturbations in the bulk, which will then backreact
on the brane through the Weyl's fluid perturbations leading to
``gravitational leakage'' into the extra dimension. The time scale for
this leakage to occurs is however of order $r_c$ which is much larger
than the age of the universe at recombination\footnote{This
qualitative picture is supported by numerical calculations in a scalar
field toy model \cite{Arthur}.}.

This discussion indicates that one can consistently use the
usual 4D cosmological perturbations equations for dealing with the
growth of small scale fluctuations observed in CMB.  The effects of
gravitational leakage is then only contained in the background
evolution, which affects the growth of the perturbations, but also the
way they appear on the sky through a different angular diameter
distance.  We only expect possible deviations on large scale coming
from the effects mentioned above, and also possible modifications once
compared with large scale structure data. We let these interesting
questions for future investigations, as well as a more careful check
of the approximations made here.

\section{CMB Likelihood calculation}
\label{appB}
Here we describe the details of the CMB likelihood calculation.

To accelerate the calculation of the model predictions we used the
k-split approximation described in \cite{Tegmark:2001qy}. In this
approximation the high $l$ power spectra is calculated in a flat model
with no dark energy and then shifted appropriately in $l$ using the
angular diameter distance to recombination. The likelihood for each
model was calculated using the RADPACK\footnote{RADPACK is publicly
available software package developed by Lloyd Knox. It can be obtained
from http://bubba.ucdavis.edu/ $\tilde{}$ knox/radpack.html} package.
We used all currently available CMB data (for a description of the
compilation we refer the reader to the RADPACK documentation and to
\cite{WangTegZal}).

For the present study we did not use a grid based method for
calculating likelihoods (such as the one described in
\cite{Tegmark:2001qy}). Following \cite{CMKL} we instead chose to use
the Metropolis Hastings algorithm to generate a Markov chain of
models.  Because we are not interested in investigating multiple
priors, our parameter space is rather small so we do not need to
exploit the CMB degeneracies and we don't want to build a database of
models to be used in future studies, the Markov-chain technique was
very efficient and extremely easy to implement.

In the Metropolis Hastings algorithm a chain of models is generated.
Models are added to the chain sequentially. To find a new model for
the chain values of the parameters are chosen at random (we choose to
select models from a Gaussian distribution centered in the last model
of the chain with a covariance matrix that is estimated from the chain
itself). The likelihood of this new model is compared to the
likelihood of the last model in the chain. The new model is always
accepted into the chain if the likelihood is larger than that of the
last model, if this is not the case it will be accepted with a
probability given by the likelihood ratio of the two models. When the
chain has converged ({\it i.e.} it has run for a sufficiently long time)
each model can be taken as an independent sample from the probability
distribution $P(\th | \d)$, the probability of some particular value
of the parameters ($\th$) given the observed data ($\d $).

Once we have the chain of models, using histograms we can construct
the distribution function of individual parameters. 

\end{document}